\documentclass[reqno,12pt]{amsart}

\usepackage{amsmath,amssymb,amsthm,amsfonts,amsxtra}
\usepackage{fullpage}

\numberwithin{equation}{section} 

{\bf}{\em}
{\bf}{\em}
{\bf}{\em}
{\bf}{\em}
{\bf}{\em}
{\bf}{\em}

\def\pe{\perp}
\def\pa{\parallel}

\def\mb{\mathbf}
\def\mr{\mathrm}
\def\mk{\mathfrak}
\def\bs{\boldsymbol}

\def\bpm{\begin{pmatrix}}
\def\epm{\end{pmatrix}}

\def\Rnum{\mathbb{R}}
\def\Cnum{\mathbb{C}}

\def\euc{{\mk{e}}}

\DeclareMathOperator{\re}{Re}
\DeclareMathOperator{\im}{Im}
\DeclareMathOperator{\spn}{span}

\DeclareMathOperator{\ad}{ad}
\DeclareMathOperator{\Ad}{Ad}

\def\t{{\rm t}}

\def\id{{\rm id}}

\def\<{\langle}
\def\>{\rangle}

\def\gsp{\mk{g}}
\def\msp{\mk{m}}
\def\hsp{\mk{h}}

\def\i{\mr{i}}
\def\j{\mr{j}}
\def\k{\mr{k}}
\def\q{\mr{q}}

\def\hatJ{\mr{J}}
\def\J{\bs{\sigma}}

\def\e{\mr{e}}

\def\hook{\rfloor}

\def\u{\mr{u}}

\def\AA{\mb{A}}
\def\BB{\mb{B}}
\def\CC{\mb{C}}
\def\DD{\mb{D}}

\def\EE{\mb{E}}
\def\UU{\mb{U}}
\def\HH{\mb{H}}

\def\WW{\mb{W}}

\def\Rop{{\mathcal R}}

\def\Jop{{\mathcal J}}
\def\Hop{{\mathcal H}}
\def\Eop{{\mathcal E}}

\def\Pop{{\mathcal P}}
\def\Iop{{\mathcal I}}

\def\Ref#1{Ref.~\cite{#1}}

\def\eg/{e.g.}
\def\ie/{i.e.}

\def\scrpt#1{$\scriptstyle {#1}$}

\begin{document}

\allowdisplaybreaks[3]
\tolerance =99999

\title{Integrable spinor/quaternion generalizations\\ of the nonlinear Schr\"{o}dinger equation}
%\\ from geometric curve flows in $SO(2N)/U(N)$}

\author{
Stephen C. Anco$^1$
\lowercase{and}
Ahmed M.G. Ahmed${}^{2}$
\lowercase{and}
Esmaeel Asadi$^3$
\\
\\
\lowercase{\scshape{
${}^1$
Department of Mathematics and Statistics\\
Brock University\\
St. Catharines, ON L\scrpt2S\scrpt3A\scrpt1, Canada\\
}}
\\
\lowercase{\scshape{
${}^2$
Department of Mathematics\\
University of South Florida\\
Tampa, FL 33620, USA\\
}}
\\
\lowercase{\scshape{
${}^3$
Department of Mathematics\\
Institute for Advance Studies in Basic Science (IASBS)\\
45137--66731, Zanjan, Iran\\
}}
}

\thanks{$^1$sanco@brocku.ca, $^2$ahmedmgahmed@hotmail.com, $^3$esmaeel.asadi@gmail.com}

\begin{abstract}
An integrable generalization of the NLS equation is presented, 
in which the dynamical complex variable $u(t,x)$ is replaced by a pair of dynamical complex variables $(u_1(t,x),u_2(t,x))$,
and $\i$ is replaced by a Pauli matrix $\J$.
Integrability is retained by the addition of a nonlocal term in the resulting 2-component system. 
A further integrable generalization is obtained 
which involves a dynamical scalar variable and an additional nonlocal term. 
For each system, a Lax pair and a bi-Hamiltonian formulation are derived from 
a zero-curvature framework that is based on symmetric Lie algebras 
and that uses Hasimoto variables. 
The systems are each shown to be equivalent to a bi-normal flow and a Schrodinger map equation, 
generalizing the well-known equivalence of the NLS equation 
to the bi-normal flow in $\Rnum^3$ and the Schrodinger map equation in $S^2$.  
Furthermore, 
both of the integrable systems describe spinor/quaternion NLS-type equations
with the pair $(u_1(t,x),u_2(t,x))$ being viewed as a spinor variable or equivalently a quaternion variable. 
\end{abstract}

\maketitle

%\tableofcontents
\date{\today,\version}

\section{Introduction}

The nonlinear Schr\"odinger (NLS) equation $u_t = \i(u_{xx} + \tfrac{1}{2}|u|^2u)$
is an important integrable system that arises in numerous areas of mathematical physics. 
As a consequence, 
there has been considerable mathematical and physical interest in finding and studying 
multi-component generalizations that retain its integrability features. 

One key feature is that the NLS equation possesses a hierarchy of symmetries
which are generated by a recursion operator applied to a root symmetry 
given by an infinitesimal phase rotation $u\to \i u$. 
These symmetries correspond to commuting bi-Hamiltonian flows 
which come from a Lax pair formulated as a zero-curvature equation 
in the Lie algebra $\mk{sl}(2,\Cnum)$. 

Here, a novel generalization of the NLS equation and its symmetry structure 
will be introduced, 
based on replacing the $U(1)$ phase rotation group by the larger group $SU(2)$, 
and replacing the complex scalar variable $u$ by a pair of complex scalar variables $(u_1,u_2)$,
with the root symmetry consisting of $(u_1,u_2)\to (u_1,u_2)\J$
where $\J$ is a Pauli matrix. 
The generalization is constructed by use of a zero-curvature equation 
in larger Lie algebras. 
This framework yields a Lax pair and a pair of compatible Hamiltonian operators 
as well as a recursion operator. 

One novelty of this generalization is that it does not retain any $U(1)$ symmetry. 
In contrast, the known multi-component generalizations of the NLS equation,
such as the Fordy-Kullish NLS systems \cite{ForKul} 
associated to Hermitian symmetric Lie algebras 
and the unitarily-invariant NLS systems \cite{AhmAncAsa2018}
associated to Riemannian symmetric Lie algebras with a unitary subalgebra, 
all involve an invariance group whose center is a $U(1)$ subgroup. 
Their root symmetry comes from multiplication by $\i$ on their variables. 
Generalizing $U(1)$ to $SU(2)$, with $\i$ replaced by $\J$, 
leads to the interesting structure that 
the variables in the resulting $SU(2)$ integrable system 
can be identified with the components of a spinor representation of $Spin(3)$, 
so thus the system will describe an integrable NLS spinor equation
for $u=(u_1,u_2)$. 

Another interesting feature of this generalization is that 
the spinor variable $u$ can be identified with a quaternion variable $\u = u_1+u_2\j$,
with $\i,\j,\k=\i\j$ being the imaginary basis quaternions. 
Similarly, $\J$ can be identified with an imaginary unit quaternion $\q$. 
In terms of the quaternion variable $\u$, 
the NLS spinor equation can be thereby expressed as a quaternion NLS equation, 
which provides a novel integrable generalization of the NLS equation 
in which $\i$ is replaced by $\q$. 

There are several zero-curvature equations from which the NLS equation can be derived.
The one that will be most useful for the present work is based on 
the Lie algebra of Euclidean space, $\euc(3) = \Rnum^3\rtimes \mk{so}(3)$.
This Lie algebra can be viewed in a natural way 
as a contraction of the symmetric Lie algebra $\mk{so}(4)$. 
It also is the one that underlies the Hasimoto transformation \cite{Has}
showing the equivalence of the NLS equation and the vortex filament equation in fluid mechanics. 

For generalizing the NLS equation to an $SU(2)$ system, 
the Euclidean Lie algebra will be replaced by a Lie algebra obtained from 
a similar contraction of the symmetric Lie algebra $\mk{su}(4)/\mk{sp}(2)$. 
Another integrable $SU(2)$ generalization will be derived by considering 
a contraction of the symmetric Lie algebra $\mk{so}(6)/\mk{u}(3)$,
which will lead to an NLS-type system involving 
a real scalar variable in addition to a pair of complex scalar variables. 
An important ingredient in these derivations will be the use of Hasimoto variables
associated naturally to the symmetric structure of the Lie algebras, 
which utilizes the general results in \Ref{Anc2008} 
yielding explicit bi-Hamiltonian operators associated to Riemannian symmetric Lie algebras. 

Through the Hasimoto transformation, 
the NLS equation is well known to correspond to a bi-normal equation 
for the motion of an inelastic (non-stretching) curve in Euclidean space. 
In turn, this motion is well know to be equivalent to a Schrodinger map into the 2-sphere, $S^2\subset\Rnum^3$. 
(For an overview, see \Ref{AncMyr,MarSanWan}.)
Similar geometric formulations hold for the two $SU(2)$ NLS-type systems. 
As a result, 
novel geometric generalizations of the bi-normal equation and the Schrodinger map 
are obtained. 

In section~\ref{sec:nls},
the zero-curvature derivation of the NLS equation from the Lie algebra $\mk{so}(4)$ 
is summarized to illustrate the method and the notation. 
Its relationship to the $\mk{sl}(2,\Cnum)$ zero-curvature framework 
will be discussed briefly. 
In section~\ref{sec:su4-sp2}, 
the derivation of the first $SU(2)$ NLS-type integrable system is presented,
and it is formulated as a quaternion equation. 
Its Lax pair and bi-Hamiltonian structure also are stated. 
In section~\ref{sec:so6-su3}, 
the second $SU(2)$ NLS-type integrable system 
and its corresponding quaternion formulation are presented, 
along with its Lax pair and bi-Hamiltonian structure. 
In section~\ref{sec:geomflows}, 
the geometric formulations of the systems are summarized. 

Some concluding remarks are made in section~\ref{sec:remarks}.

\section{Zero-curvature derivation of the NLS equation}\label{sec:nls}

To begin, some brief general preliminaries will be summarized. 
(For more details, see \Ref{Hel,Anc2008}.)
A symmetric Lie algebra, $(\gsp,\hsp)$, 
has the structure $\gsp=\msp\oplus\hsp$
in which $\hsp$ is a subalgebra and $\msp$ is a subspace 
with the bracket relations 
\begin{equation}\label{brack.rel}
[\hsp,\hsp]\subset \hsp,
\quad
[\hsp,\msp] \subseteq \msp,
\quad 
[\msp,\msp]\subset \hsp . 
\end{equation}
There is Cartan-Killing form $K$ on $\gsp$,
where $K(\hsp,\msp) =0$. 
When $\gsp$ is compact, 
the Cartan-Killing form provides a negative-definite inner product on $\msp$. 

Contraction of $\gsp$ consists of scaling $\msp\to\epsilon\msp$ 
and taking the singular limit $\epsilon\to0$,
which changes the third bracket relation: $[\msp,\msp]\to 0$. 
This yields a Lie algebra given by the semi-direct product structure $\msp\rtimes\hsp$,
retaining the first and second bracket relations \eqref{brack.rel},
with $\msp$ becoming an abelian subalgebra.

\subsection{NLS equation from $\mk{so}(4)/\mk{so}(3)$}

Consider the symmetric Lie algebra $(\mk{so}(4),\mk{so}(3))$,
which has $\msp = \Rnum^3 \simeq \mk{so}(4)/\mk{so}(3)$ and $\hsp = \mk{so}(3)$. 

The zero-curvature framework in this Lie algebra is set up as follows. 
Choose a constant element $\e$ belonging to a Cartan subspace in $\msp$, 
and let $H_{\pa}\subset SO(3)$ denote the linear isotropy group that preserves $\e$,
namely $\Ad(H_\pa)\e = \e$. 
The corresponding Lie subalgebra will be denoted $\hsp_{\pa}\subset \hsp$, 
where $\ad(\hsp_{\pa})\e=0$. 
There is an orthogonal decomposition $\hsp=\hsp_{\pa}\oplus\hsp_{\perp}$, 
where $K(\hsp_{\pa},\hsp_{\perp})=0$. 
Similarly, there is an orthogonal decomposition $\msp=\msp_{\pa}\oplus\msp_{\perp}$,
where $\ad(\msp_{\pa})\e=0$ 
and $K(\msp_{\pa},\msp_{\perp})=0$. 

A matrix representation of this structure is given by 
\begin{gather}
\msp = \bpm 0 & a_1 & a_2 & a_3 \\ -a_1 & 0 & 0 & 0 \\ -a_2 & 0 & 0 & 0 \\ -a_3 & 0 & 0 & 0 \epm \simeq \Rnum^3 ,
\quad
\hsp = \bpm 0 & 0 & 0 & 0 \\ 0 & 0 & b_1 & b_2 \\ 0 & -b_1 & 0 & b_3 \\ 0 & -b_2 & -b_3 & 0 \epm \simeq \mk{so}(3)
\end{gather}
and 
\begin{equation}
\e = \chi\bpm 0 & 1 & 0 & 0 \\ -1 & 0 & 0 & 0 \\ 0 & 0 & 0 & 0 \\ 0 & 0 & 0 & 0 \epm 
\in \msp . 
\end{equation}
Here $\chi$ is arbitrary non-zero constant related to norm of $\e$:
$K(\e,\e)=-2\chi^2$. 
Note that, because $(\mk{so}(4),\mk{so}(3))$ has rank $1$ \cite{Hel}, 
its Cartan subspaces have dimension $1$ and are equivalent to $\spn(\e)$ 
up to the action of $\Ad(SO(3))$. 

The orthogonal decompositions are represented by 
\begin{gather}
\msp_\pa = \bpm 0 & a_1 & 0 & 0 \\ -a_1 & 0 & 0 & 0 \\ 0 & 0 & 0 & 0 \\ 0 & 0 & 0 & 0 \epm 
\simeq \Rnum,
\quad
\msp_\pe = \bpm 0 & 0 & a_2 & a_3 \\ 0 & 0 & 0 & 0 \\ -a_2 & 0 & 0 & 0 \\ -a_3 & 0 & 0 & 0 \epm 
\simeq \Rnum\times\Rnum\simeq \Cnum,
\\
\hsp_\pa = \bpm 0 & 0 & 0 & 0 \\ 0 & 0 & 0 & 0 \\ 0 & 0 & 0 & b_3 \\ 0 & 0 & -b_3 & 0 \epm \simeq \mk{so}(2)=\mk{u}(1),
\quad
\hsp_\pe = \bpm 0 & 0 & 0 & 0 \\ 0 & 0 & b_1 & b_2 \\ 0 & -b_1 & 0 & 0 \\ 0 & -b_2 & 0 & 0 \epm 
\simeq \Rnum\times\Rnum\simeq \Cnum . 
\end{gather}
It will be useful to consider the scaled space 
\begin{equation}
\hsp^\pe :=\tfrac{1}{\chi^2}\ad(\e)\msp_\pe
= -\tfrac{1}{\chi}\bpm 0 & 0 & 0 & 0 \\ 0 & 0 & b_1 & b_2 \\ 0 & -b_1 & 0 & 0 \\ 0 & -b_2 & 0 & 0 \epm 
\simeq \hsp_\pe .
\end{equation}

Associated to $\hsp^\pe$ will be the flow variables 
$(h_1^\pe,h_2^\pe)\in\Rnum\times\Rnum\simeq\Cnum$. 
The Hasimoto variables \cite{Anc2008} are defined by belonging to $\hsp_\pe$: 
$(u_1,u_2)\in\Rnum\times\Rnum\simeq\Cnum$. 
Under the action of $\mk{so}(2)$, 
the Hasimoto variables undergo a rotation generated by 
$(u_1,u_2)\to (-u_2,u_1)$. 
Then the zero-curvature equation in $\mk{so}(4)$ is given by 
the matrices
\begin{equation}\label{nls.U.V}
U=\bpm 0 & \chi & 0 & 0 \\ -\chi & 0 & u_1 & u_2 \\ 0 & -u_1 & 0 & 0 \\ 0 & -u_2 & 0 & 0\epm
\in\msp_\pa\oplus\hsp_\pe, 
\quad
V=\bpm 0 & \chi h_\pa & -\chi h^{1\pe} & -\chi h^{2\pe} \\ -\chi h_\pa & 0 & w_{1\pe} & w_{2\pe} \\ \chi h^{1\pe} & -w_{1\pe} & 0 & w_\pa \\ \chi h^{2\pe} & -w_{2\pe} & -w_\pa & 0 \epm
\in \msp\oplus\hsp
\end{equation}
%$(h_1^\pe,h_2^\pe) =-\tfrac{1}{\chi}(h_{1\pe},h_{2\pe})$
which are taken to satisfy 
\begin{equation}\label{zerocurveqn}
D_t U -D_x V  -[U,V]=0 . 
\end{equation}
The components of this matrix equation yield the system 
\begin{subequations}\label{nls.zerocurv.system}
\begin{align}
& 
D_x h_\pa = u_1 h_1^\pe + u_2 h_2^\pe,
\quad
D_x w_\pa = u_1 w_{2\pe} - u_2 w_{1\pe} ,
\\
& 
w_{1\pe} = h_\pa u_1 + D_x h_1^\pe,
\quad
w_{2\pe} = h_\pa u_2 + D_x h_2^\pe ,
\\
&
u_1{}_t = D_x w_{1\pe} -w_\pa u_2 +\chi^2 h_1^\pe,
\quad
u_2{}_t = D_x w_{2\pe} +w_\pa u_1 +\chi^2 h_2^\pe .
\end{align}
\end{subequations}
The isotropy group $H_\pa = SO(2)$ acts as a 
rotation on $(u_1,u_2)$, $(w_{1\pe},w_{2\pe})$, $(h_1^\pe,h_2^\pe)$. 
Hence, it is natural to go to complex variables
\begin{equation}\label{so4.complexvars}
u = u_1+\i u_2,
\quad
w = w_{1\pe}+\i w_{2\pe},
\quad
h^\pe = h_1^\pe +\i h_2^\pe
\end{equation}
on which the isotropy group acts a $U(1)$ phase rotation. 
In particular, this is a symmetry group of the system \eqref{nls.zerocurv.system}. 

In terms of the complex variables \eqref{so4.complexvars}, the system has the form 
\begin{subequations}\label{nls.zerocurv.complexvars.system}
\begin{align}
& 
D_x h_\pa = \re(\bar u h^\pe) , 
\\
& 
w = h_\pa u + D_x h^\pe , 
\\
& 
D_x w_\pa = \im(\bar u w) , 
\\
& 
u_t = D_x w +\i w_\pa u +\chi^2 h^\pe . 
\end{align}
\end{subequations}

The action of the $U(1)$ symmetry consists of a phase rotation $e^{\i\phi}$. 
Its generator is now used to define a flow (see \Ref{Anc2008,AhmAncAsa2018})
by putting
\begin{equation}
h^\pe = \i u .
\end{equation}
The system \eqref{nls.zerocurv.complexvars.system} thereby yields 
$h_\pa = c_1$, $w=\i u_x +c_1 u$, $w_\pa = \tfrac{1}{2} u \bar u +c_2$,
and 
\begin{equation}
u_t = \i( u_{xx} + \tfrac{1}{2} |u|^2 u ) +c_1 u_x + (c_2 +\chi^2)\i u
\end{equation}
where $c_1,c_2$ are arbitrary constants. 
Thus, the NLS equation is obtained for $c_1=0$ and $c_2=-\chi^2$. 

The integrability structure of the NLS equation is encoded directly in 
the zero-curvature system \eqref{nls.zerocurv.complexvars.system}. 

First, through elimination of the variables $h_\pa$ and $w_\pa$, 
the system can be expressed in the operator form 
\begin{equation}\label{Hamil.form}
u_t = \Hop(w),
\quad 
w = \Jop(h^\pe)
\end{equation}
with 
$\Hop=D_x +\i u D_x^{-1} \im\bar u$ 
and
$\Jop= D_x + u D_x^{-1}\re\bar u$. 
The general theorem in \Ref{Anc2008} 
establishes that $\Hop$ is a Hamiltonian operator
and $\Jop$ is a compatible symplectic operator,
which is well known. 
Their composition $\Rop=\Hop\Jop$ is a symmetry recursion operator. 
This yields the Hamiltonian formulation 
$u_t= \Hop(\delta\mk{H}/\delta\bar u) = \Eop(\delta\mk{E}/\delta\bar u)$, 
where $\Eop= \Rop\Hop$ is Hamiltonian operator compatible with $\Hop$. 
The first Hamiltonian is given by 
$\mk{H}=\int \im(\bar u_x u)\,dx$; 
the second Hamiltonian is $\mk{E}=0$, using $D_x^{-1}(0)=c=$constant, 
with $c=1$. 

Second, the matrices \eqref{nls.U.V} and the zero-curvature equation \eqref{zerocurveqn}
define a matrix Lax pair, 
with $h^\pe = \i u$, $h_\pa = 0$, $w=\i u_x$, $w_\pa = \tfrac{1}{2} u \bar u -\chi^2$,
as given by solving the zero-curvature system \eqref{nls.zerocurv.complexvars.system}:
\begin{align}
& 
U=\bpm 0 & \chi & 0 & 0 \\ -\chi & 0 & \re u & \im u \\ 0 & -\re u & 0 & 0 \\ 0 & -\im u & 0 & 0\epm,
\quad
V=\bpm 0 & 0  & \chi \im u & -\chi \re u \\ 0 & 0 & -\im u_x & -\re u_x \\ -\chi \im u & \im u_x  & 0 & \tfrac{1}{2}|u|^2 -\chi^2 \\ \chi \re u & \re u_x & -\tfrac{1}{2}|u|^2 +\chi^2 & 0 \epm . 
\end{align}
In this Lax pair, $\chi$ represents the spectral parameter. 

It is straightforward to convert this Lax pair into a matrix representation 
for the contracted Lie algebra $\Rnum^3\rtimes\mk{so}(3)$ 
which is the Lie algebra of the isometry group $ISO(3)$ of $\Rnum^3$. 
At the level of the zero-curvature system \eqref{nls.zerocurv.complexvars.system}, 
the contraction simply removes the term $\chi^2 h^\pe$ 
in the flow equation for the Hasimoto variable $u$. 

There is an alternative zero-curvature equation that yields the NLS equation. 
It uses the Lie algebra $\mk{so}(3,1)\simeq \mk{sl}(2,\Cnum)$,
which is a symmetric Lie algebra 
where $\msp = \Rnum^{2,1}$ is $3$-dimensional Minkowski space
and $\hsp=\mk{so}(2,1)\simeq \mk{sl}(2,\Rnum)$ is the Lie algebra of rotations and boosts.  
The only change in the previous construction is that $\e$ is chosen to be a constant element 
corresponding to a timelike vector in $\Rnum^{2,1}$. 
The resulting matrices $U$, $V$ are the directly related to the well-known 
AKNS scheme \cite{AblKauNewSeg} for the NLS equation.

\section{First NLS-type system and its integrability}\label{sec:su4-sp2}

For writing down the $SU(2)$ generalization of the NLS equation, 
take a general element in the Lie algebra $\mk{su}(2)$ 
and normalize it to get 
\begin{equation}\label{hatJ}
\hatJ := \bpm \i\cos\theta & e^{\i\psi}\sin\theta \\ -e^{-\i\psi}\sin\theta & -\i\cos\theta \epm
\in\mk{su}(2),
\quad
\theta,\psi\in[0,2\pi)
\end{equation}
(with $|\hatJ|=1$ being the absolute norm given by the Cartan-Killing metric). 
This normalized general element has the properties
\begin{equation}
\bar\hatJ^\t = -\hatJ,
\quad
\hatJ^2=-\id,
\quad
\det(\exp(\phi\hatJ))=1
\end{equation}
which describe an $SU(2)$ analog of the element $\i$ in $\mk{u}(1)$:
$\bar\i=-\i$, $\i^2=-1$, $|e^{\i\phi}|=1$. 
(Throughout, 
a superscript ``$\t$'' denotes the transpose;
a subscript ``$0$'' denotes trace-free part of a matrix.)

The $SU(2)$ generalization of the NLS equation has the form 
\begin{equation}\label{1st.su2.ueqn}
u_t =
u_{xx}\hatJ  
+|u|^2 u\hatJ 
+2 u D_x^{-1}( \bar u^\t u_x \hatJ + \hatJ \bar u_x^\t u )_0
\end{equation}
in terms of the variable 
\begin{equation}\label{u.pair}
u:= \bpm u_1 & u_2 \epm \in \Cnum\times\Cnum=\Cnum^2,
\quad
u_1,u_2\in\Cnum . 
\end{equation}

The nonlocal $D_x^{-1}$ term can be rewritten in various ways via the identities
\begin{equation}\label{u.identities}
u (\hatJ\bar u^\t u)_0 = \tfrac{1}{2}\i \im(\bar u \cdot(u \hatJ)) u ,
\quad
u (\bar u^\t u \hatJ)_0 = |u|^2 u\hatJ - \tfrac{1}{2}\i \im(\bar u \cdot(u \hatJ)) u .
\end{equation}
In particular: 
\begin{equation}\label{1st.su2.ueqn.alt}
\begin{aligned}
u_t - u_{xx}\hatJ  & 
= |u|^2 u\hatJ +\tfrac{1}{4}\i \im(\bar u \cdot(u \hatJ)) u
+2 u D_x^{-1}[\bar u^\t u_x, \hatJ]
\\& 
=  3 |u|^2 u\hatJ -\i \im(\bar u \cdot(u \hatJ)) u
-2 u D_x^{-1}[\bar u_x^\t u,\hatJ]
\\& 
= 2 |u|^2 u\hatJ 
+2 u D_x^{-1}[\bar u^\t u_x-\bar u_x^\t u,\hatJ] .
\end{aligned}
\end{equation}

Since $u$ naturally represents a spinor variable, 
the equation \eqref{1st.su2.ueqn} describes a spinor generalization of the NLS equation. 
Moreover, the identifications
\begin{equation}\label{u.quaternion}
u = (u_1,u_2) \leftrightarrow \u = u_1+u_2 \j,
\quad
\hatJ \leftrightarrow \q
\end{equation}
allows the equation \eqref{1st.su2.ueqn} to be expressed as a quaternion NLS equation:
\begin{equation}
\u_t = \u_{xx}\q +|\u|^2 \u\q
+\u D_x^{-1}( \bar\u \u_x \q + \q \bar\u_x \u )
\end{equation}
where $\q$ is an imaginary unit quaternion
\begin{equation}\label{q}
\q^2=-1, 
\quad
\q\bar\q =1 .
\end{equation}

\subsection{Integrability structure}

The $SU(2)$ NLS-type equation \eqref{1st.su2.ueqn}
is an integrable system. 

First, it possesses a symmetry recursion operator 
$\Rop=\Hop\Jop$ 
where 
\begin{equation}\label{1st.Hop}
\Hop  =D_x +\i u D_x^{-1}\im \bar u\cdot\ + 2u D_x^{-1}\Pop_{\mk{su}}\bar u^\t\ +4\bar u D_x^{-1}\Pop_{\mk{so}} u^\t\
\end{equation}
is a Hamiltonian operator,
and where
\begin{equation}\label{1st.Jop}
\Jop = D_x - 4 u D_x^{-1}\re \bar u\cdot\
\end{equation}
is a symplectic operator. 
Here $\Pop_{\mk{su}}$ is a projection onto the skew-Hermitian, trace-free part of a matrix;
$\Pop_{\mk{so}}$ is a projection onto the skew part of a matrix. 

Second, this pair of operators \eqref{1st.Hop}--\eqref{1st.Jop}
provides a bi-Hamiltonian formulation
\begin{equation}
u_t= \Hop(\delta\mk{H}/\delta\bar u^\t) = \Eop(\delta\mk{E}/\delta\bar u^\t)
\end{equation}
where $\Eop= \Rop\Hop$ is Hamiltonian operator compatible with $\Hop$. 
The first Hamiltonian is given by 
$\mk{H}=\int \re(\bar u\cdot (u_xJ))\,dx$; 
the second Hamiltonian is $\mk{E}=0$ with 
$D_x^{-1}(0)=0$, $D_x^{-1}(\mb{0}) = \hatJ$
in $\Hop$. 
An interesting remark is that 
the $\mk{su}(2)$ generator $\hatJ$ is also a Hamiltonian operator, 
however, 
the $SU(2)$ equation \eqref{1st.su2.ueqn} is not Hamiltonian 
with respect to this operator. 
This stands in contrast to the structure of the NLS equation, 
where $\i$ is a Hamiltonian operator that is compatible with the other two. 

Third, the $SU(2)$ equation \eqref{1st.su2.ueqn} possess a Lax pair \eqref{zerocurveqn}
given by the $\mk{su}(4)$ matrices 
\begin{gather}
U=
\bpm 
\i\chi &  u_1 & 0 & \bar u_2 \\ 
-\bar u_1 & -\i\chi & \bar u_2 & 0\\ 
0 & -u_2 & \i\chi & \bar u_1 \\ 
-u_2 & 0 & -u_1 & -\i\chi
\epm ,
\\
\begin{aligned}
& V= \\& 
\bpm 
\i w_{1\pa} & -2\i\chi (u\hatJ)_1 + 4 (u_x\hatJ)_1 & w_{2\pa} & -2\i\chi (\bar u \bar\hatJ)_2 +4(\bar u_x\bar\hatJ)_2 \\ 
-2\i\chi (\bar u\bar\hatJ)_1 -4(\bar u_x\bar\hatJ)_1 & \bar u\cdot(u\hatJ) & 2\i\chi(\bar u\bar\hatJ)_2 +4(\bar u_x\bar\hatJ)_2  & W_{2\pa}\\
-\bar w_{2\pa} & 2\i\chi (u\hatJ)_2 -4(u_x \hatJ)_2 & -\i w_{1\pa}  & -2\i\chi(\bar u\bar\hatJ)_1 + 4(\bar u_x\bar\hatJ)_1 \\
-2\i\chi(u\hatJ)_2  -4(u_x\hatJ)_2 & -\bar W_{2\pa} & -2\i\chi(u\hatJ)_1 -4(u_x \hatJ)_1 &  -\bar u\cdot(u\hatJ) 
\epm 
\end{aligned}
\end{gather}
with 
\begin{equation}
\bpm 0 & \bar W_{2\pa} \\ -\bar W_{2\pa} & 0 \epm
= u^\t u\hatJ - \hatJ^\t u^\t u ,
\quad
\bpm -\i w_{1\pa} & -\bar w_{2\pa} \\ w_{2\pa} & \i w_{1\pa} \epm
= 2 D_x^{-1}( \bar u^\t  u_x \hatJ + \hatJ \bar u_x^\t u ) -\chi^2\hatJ .
\end{equation}
The spectral parameter in this Lax pair is $\chi$.

\subsection{Derivation}

The $SU(2)$ equation \eqref{1st.su2.ueqn} comes from 
the general zero-curvature framework outlined in section~\ref{sec:nls},
using the symmetric Lie algebra 
$(\mk{su}(4),\mk{sp}(2))\simeq (\mk{so}(6),\mk{so}(5))$. 
This Lie algebra has the bracket relations \eqref{brack.rel}
with $\hsp = \mk{sp}(2)$, $\msp = \Rnum^5 \simeq \mk{su}(4)/\mk{sp}(2)$,
and $\gsp=\mk{su}(4)$. 
There is a unique choice of $\e$ in a Cartan space in $\msp$
up to gauge freedom generated by $\ad(\hsp)$,
since the rank of the Lie algebra is $1$ \cite{Hel}. 

A matrix representation is given by 
\begin{align}
\hsp & = \bpm \CC & \DD \\ -\bar\DD & \bar\CC \epm 
\simeq \mk{sp}(2), 
\quad 
\CC\in\mk{u}(2), \DD\in\mk{s}(2,\Cnum) 
\label{su.hsp.matr.rep}
\\
\msp & = \bpm \AA & \BB \\ \bar\BB & -\bar\AA \epm 
\simeq \Rnum^5,
\quad 
\AA\in\mk{su}(2), \BB\in\mk{so}(2,\Cnum)
\label{su.msp.matr.rep}
\end{align}
and
\begin{equation}
\e = \bpm \EE & \mb{0} \\ \mb{0} & \EE \epm
\in\msp,
\quad
\EE = 
\chi\bpm \i & 0 \\ 0 & -\i \epm
\in\mk{su}(2) 
\end{equation}
where $\chi$ is an arbitrary non-zero real constant related to norm of $\e$
(namely, $K(\e,\e)=-4\chi^2$). 
The isotropy group preserving $\e$ is $H_\pa = SU(2)\times Sp(1)\subset Sp(2)$,
whose Lie algebra is $\hsp_\pa\simeq\mk{su}(2)\oplus\mk{sp}(1)\subset\mk{sp}(2)$. 
This Lie algebra contains the normalized generator \eqref{hatJ}:
\begin{equation}\label{su.symm.su2}
\hatJ \simeq \bpm \CC_\hatJ & \DD_\hatJ \\ -\bar\DD_\hatJ & \bar\CC_\hatJ \epm , 
\quad
\CC_\hatJ = 
\bpm \i\cos\theta  & 0 \\ 0 & -\i\cos\theta \epm , 
\quad
\DD_\hatJ = \bpm e^{-\i\psi}\sin\theta & 0 \\ 0 & 0 \epm . 
\end{equation}

The orthogonal decomposition $\msp=\msp_\pa\oplus\msp_\pe$ 
corresponds to $\AA = \AA_\pa+\AA_\pe$ and $\BB = \BB_\pa+\BB_\pe$: 
\begin{align}
\AA_\pa &= 
\bpm -\i a_\pa & 0\\ 0 & \i a_\pa \epm
\in\mk{su}(2),
\quad
\BB_\pa = \mb{0},
\quad
a_\pa\in\Rnum
\label{su.msp.par.matr.rep}
\\
\AA_\pe &= 
\bpm 0& a_{\pe} \\  -\bar a_\pe & 0 \epm
\in\mk{su}(2),
\quad
\BB_\pe = 
\bpm 0&  b_{\pe} \\  - b_\pe & 0 \epm
\in\mk{so}(2,\Cnum),
\quad
a_\pe, b_\pe\in\Cnum . 
\label{su.msp.perp.matr.rep}
\end{align}
Likewise, 
the orthogonal decomposition $\hsp=\hsp_\pa\oplus\hsp_\pe$ 
corresponds to $\CC = \CC_\pa+\CC_\pe$ and $\DD = \DD_\pa+\DD_\pe$:
\begin{align}
\CC_\pa &= 
\bpm \i c_\pa  & 0 \\ 0 & \i C_\pa \epm
\in\mk{u}(2),
\quad
\DD_\pa = \bpm  d_\pa  & 0 \\ 0 &  D_\pa \epm
\in\mk{s}(2,\Cnum),
\quad
 c_\pa, C_\pa\in\Rnum,
 d_\pa, D_\pa\in\Cnum
\label{su.hsp.par.matr.rep}
\\
\CC_\pe &= 
\bpm 0 &  c_\pe\\ -\bar c_\pe & 0 \epm
\in\mk{u}(2),
\quad
\DD_\pe = 
\bpm 0 &  d_\pe\\  d_\pe & 0 \epm
\in\mk{s}(2,\Cnum),
\quad
 c_\pe, d_\pe\in \Cnum . 
\label{su.hsp.perp.matr.rep}
\end{align}
The scaled space $\hsp^\pe:=\tfrac{1}{\chi^2}\ad(\e)\msp_\pe\simeq\hsp_\pe$
is given by the correspondence
\begin{equation}
\CC^\pe := \tfrac{2}{\chi}
\bpm 0 & \i a_\pe\\ \i\bar a_\pe & 0 \epm , 
\quad
\DD^\pe := \tfrac{2}{\chi}
\bpm 0 & \i b_\pe\\ \i b_\pe & 0 \epm . 
\end{equation}

The matrices in the zero-curvature equation \eqref{zerocurveqn} in $\mk{su}(4)$ 
are given by 
\begin{equation}\label{su.U.V}
U =
\bpm \EE+\UU_1 & \UU_2 \\ -\bar\UU_2 & \EE +\bar\UU_1 \epm , 
\quad
V =
\bpm \HH_1+\WW_1 & \HH_2+\WW_2 \\ \bar\HH_2-\bar\WW_2 & -\bar\HH_1 + \bar\WW_1 \epm , 
\end{equation}
where
\begin{align}
& 
\UU_1  = \bpm 0 & u_1\\ -\bar u_1 & 0 \epm,
\quad
\UU_2 = \bpm 0 & \bar u_2 \\ \bar u_2 & 0 \epm,
\quad
u_1,u_2\in\Cnum
\label{su.UU}
\\&
\WW_1  = \bpm \i w_{1\pa} & w_{1\pe}\\ -\bar w_{1\pe} & \i W_{1\pa} \epm,
\quad
\WW_2 = \bpm w_{2\pa} & \bar w_{2\pe} \\ \bar w_{2\pe} & W_{2\pa} \epm,
\quad
w_{1\pa},W_{1\pa}\in\Rnum,
w_{1\pe},w_{2\pe},w_{2\pa},W_{2\pa}\in\Cnum
\label{su.WW}
\\&
%\HH_1 =\bpm -\i \chi h_\pa & h_{1\pe}\\ -\bar h_{1\pe} & \i \chi h_\pa \epm
\HH_1 =\bpm -\i \chi h_\pa & -\i\tfrac{1}{2}\chi h^1_\pe\\ -\i\tfrac{1}{2}\chi \bar h_{1\pe} & \i \chi h_\pa \epm,
\quad
%\HH_2 =\bpm 0 & \bar h_{2\pe}\\ -\bar h_{2\pe} & 0 \epm
\HH_2 =\bpm 0 & \i\tfrac{1}{2}\chi \bar h_2^\pe\\  -\i\tfrac{1}{2}\chi \bar h_2^\pe& 0 \epm,
\quad
h_\pa\in\Rnum,
h_1^\pe,h_2^\pe\in\Cnum. 
\label{su.HH}
\end{align}
Here $(u_1,u_2)$ are the Hasimoto variables, 
and $(h_1^\pe,h_2^\pe)$ are the flow variables associated to the space $\hsp^\pe$
(see \Ref{Anc2008}). 
%$(h_1^\pe,h_2^\pe) =\tfrac{2}{\chi}(\i h_{1\pe},\i h_{2\pe})$
%$(h_{1\pe},h_{2\pe}) =-\tfrac{\chi}{2}(\i h_1^\pe,\i h_2^\pe)$
%$(\bar h_{1\pe},\bar h_{2\pe}) =\tfrac{\chi}{2}(\i \bar h_1^\pe,\i \bar h_2^\pe)$

The components of the zero-curvature equation yield the system 
\begin{subequations}\label{su.zerocurv.components}
\begin{align}
& 
D_x h_\pa = -\re( \bar u_1 h_1^\pe + \bar u_2 h_2^\pe ),
\\
&
D_x w_{1\pa} = -\im( \bar u_1 w_{1\pe} - \bar u_2 w_{2\pe} ),
\quad
D_x W_{1\pa} = \im( \bar u_1 w_{1\pe} + \bar u_2 w_{2\pe} ),
\\
&
D_x w_{2\pa} = 2( \bar u_2 w_{1\pe} - u_1 \bar w_{2\pe} ),
\quad
D_x W_{2\pa} = 2( \bar u_1 \bar w_{2\pe} - \bar u_2 \bar w_{1\pe} ),
\\
& 
w_{1\pe} = \tfrac{1}{4} D_x h_1^\pe -u_1 h_\pa ,
\quad
w_{2\pe} = \tfrac{1}{4} D_x h_2^\pe -u_2 h_\pa ,
\\
&
u_1{}_t = D_x w_{1\pe} +\i u_1( W_{1\pa} -w_{1\pa} ) +u_2 w_{2\pa} -\bar u_2 \bar W_{2\pa} +\chi^2 h_1^\pe , 
\\
&
u_2{}_t = D_x w_{2\pe} +\i u_2( W_{1\pa} +w_{1\pa} ) -u_1 \bar w_{2\pa} +\bar u_1 \bar W_{2\pa} +\chi^2 h_2^\pe . 
\end{align}
\end{subequations}
This system is invariant under an $SU(2)$ symmetry 
coming from the action of the isotropy group $H_\pa = SU(2)$ 
on $(u_1,u_2)$, $(w_{1\pe},w_{2\pe})$, $(h_1^\pe,h_2^\pe)$. 
The symmetry action looks simplest by going to the 2-component spinor variable \eqref{u.pair} 
along with the similar variables
\begin{equation}\label{w.hperp.pair}
w := (w_{1\pe},w_{2\pe}) \in \Cnum\times\Cnum,
\quad
h^\pe := (h_1^\pe, h_2^\pe) \in \Cnum\times\Cnum,
\end{equation}
and the matrix variables
\begin{equation}\label{su.wpar.matr}
\mb{w}_\pa := \bpm -\i w_{1\pa} & -\bar w_{2\pa} \\ w_{2\pa} & \i w_{1\pa} \epm
\in\mk{su}(2), 
\quad
\WW_\pa := \bpm 0 & \bar W_{2\pa} \\ -\bar W_{2\pa} & 0 \epm
\in\mk{so}(2,\Cnum) . 
\end{equation}
Also, for ease of notation, put $W_\pa := W_{1\pa}$. 

Thus, the system \eqref{su.zerocurv.components} takes the invariant form 
\begin{subequations}\label{su.zerocurv.system}
\begin{align}
& 
D_x h_\pa = -\re(\bar u\cdot h^\pe),
\\
& 
w = \tfrac{1}{4}D_x h^\pe + h_\pa u ,
\\
& 
D_x \WW_\pa = 2(u^\t w -w^\t u),
\quad
D_x \mb{w}_\pa = ( \bar u^\t  w - \bar w^\t u )_0,
\quad
D_x W_\pa = \im(\bar u\cdot w),
\\
& 
u_t = D_x w +\i W_\pa u + u\mb{w}_\pa +\bar u \WW_\pa +\chi^2 h^\pe
\end{align}
\end{subequations}
with the normalized $SU(2)$ symmetry generator \eqref{su.symm.su2} 
being $u\to u\hatJ$
on the Hasimoto spinor variable $u$. 

Now, the $SU(2)$ symmetry generator is used to define a flow 
\begin{equation}
h^\pe = u\hatJ
\end{equation}
The system \eqref{su.zerocurv.system} thereby yields 
\begin{gather}
h_\pa = c_1
\\
w=\tfrac{1}{4} u_x\hatJ +c_1 u
\\
\begin{gathered}
W_\pa = \tfrac{1}{4}\im(\bar u\cdot(u\hatJ))+C_1
\\
\WW_\pa = \tfrac{1}{4}( u^\t u\hatJ - (u\hatJ)^\t u ) +\CC_1,
\quad
\mb{w}_\pa = \tfrac{1}{2}D_x^{-1}( \bar u^\t u_x \hatJ + \hatJ \bar u_x^\t u )_0 +\CC_2
%= \tfrac{1}{2}D_x^{-1}( [\hatJ, \bar u_x^\t u] )_0 +\tfrac{1}{2}|u|^2 u\hatJ -\tfrac{1}{4}\i\im( \bar u (u\hatJ) ) 
\end{gathered}
\end{gather}
and 
\begin{equation}
\begin{aligned}
u_t & 
=\tfrac{1}{4}u_{xx}\hatJ  +\tfrac{1}{4}|u|^2 u\hatJ 
+ \tfrac{1}{2} u D_x^{-1}( \bar u^\t u_x \hatJ + \hatJ \bar u_x^\t u )_0
\\&\qquad
+c_1 u_x +C_1\i u+u (\CC_2 +\chi^2 \hatJ) +\bar u\CC_1 ,
\end{aligned}
\end{equation}
where $c_1,C_1$ are arbitrary constants,
$\CC_1$ is an arbitrary constant matrix in $\mk{so}(2,\Cnum)$,
and $\CC_2$ is an arbitrary constant matrix in $\mk{su}(2,\Cnum)$.
This yields the $SU(2)$ NLS-type equation \eqref{1st.su2.ueqn}
after a scaling of $t$, 
with $c_1=C_1=0$, $\CC_1=\mb{0}$, and $\CC_2=-\chi^2\hatJ$. 

The integrability structure is obtained directly from the system \eqref{su.zerocurv.system} 
by the same method shown for the NLS equation in section~\ref{sec:nls}.

\section{Second NLS-type system and its integrability}\label{sec:so6-su3}

The second $SU(2)$ generalization of the NLS equation 
has the form 
\begin{equation}\label{2nd.su2.ueqn}
\begin{aligned}
v_t & = 2\im( \bar u_x\cdot (u\hatJ)) 
\\
u_t & = 
u_{xx}\hatJ -\i v_x u\hatJ + v^2 u\hatJ 
+\tfrac{1}{2}\i \im( \bar u\cdot (u\hatJ) ) u
\\&\qquad
+ u D_x^{-1}( \bar u^\t u_x \hatJ + \hatJ \bar u_x^\t u )_0 + \i u D_x^{-1}(v [\hatJ,\bar u^\t u])
\end{aligned}
\end{equation}
where $u$ is the spinor variable \eqref{u.pair},
and $v$ is a real scalar variable. 
Through the identities \eqref{u.identities}, 
the nonlocal $D_x^{-1}$ terms can be rewritten in several ways:
\begin{equation}\label{2nd.su2.ueqn.alt}
\begin{aligned}
u_t - u_{xx}\hatJ +\i v_x u\hatJ - v^2 u\hatJ 
& =  \i \im( \bar u\cdot (u\hatJ) ) u
+ u D_x^{-1}[\bar u^\t u_x, \hatJ] + \i u D_x^{-1}(v [\hatJ,\bar u^\t u])
\\
& = |u|^2 u\hatJ - u D_x^{-1}[\bar u_x^\t u, \hatJ] + \i u D_x^{-1}(v [\hatJ,\bar u^\t u])
\\
& = \tfrac{1}{2} |u|^2 u\hatJ
+\tfrac{1}{2} \i \im( \bar u\cdot (u\hatJ) ) u
\\&\qquad
+ \tfrac{1}{2} u D_x^{-1}[\bar u^\t u_x-\bar u_x^\t u, \hatJ] + \i u D_x^{-1}(v [\hatJ,\bar u^\t u]) . 
\end{aligned}
\end{equation}

The system \eqref{2nd.su2.ueqn} describes a scalar-spinor generalization of the NLS equation. 
Through the identifications \eqref{u.quaternion}, 
it can be expressed as a quaternion NLS system:
\begin{equation}
\begin{aligned}
v_t & = 2\im( \bar\u_x \u\q  ) , 
\\
\u_t & = 
\u_{xx}\q -\i v_x \u\q + v^2 \u\q
+\tfrac{1}{4}\i [\q,\bar\u \i\u] \u
\\&\qquad
+ \u D_x^{-1}( \bar\u \u_x \q + \q \bar\u_x \u ) + \tfrac{1}{2}\u D_x^{-1}(v [\q,\bar\u\i\u])
\end{aligned}
\end{equation}
where $\q$ is an imaginary unit quaternion \eqref{q}.

\subsection{Integrability structure}

This $SU(2)$ NLS-type system \eqref{2nd.su2.ueqn}
has an integrability structure similar to that of equation \eqref{1st.su2.ueqn}. 

It possesses a symmetry recursion operator 
$\Rop=\Hop\Jop$ 
given in terms of a Hamiltonian operator
\begin{equation}\label{2nd.Hop}
\Hop =
\bpm 
D_x 
& \im \bar u\cdot\
\\
-\i u
& D_x +\i v +2\i u D_x^{-1}\im \bar u\cdot\ + u D_x^{-1}\Pop_{\mk{su}}\bar u^\t\ 
\epm
\end{equation}
and a symplectic operator
\begin{equation}\label{2nd.Jop}
\Jop = 
\bpm
\tfrac{1}{4}D_x +v D_x^{-1} v
& \tfrac{1}{2}\im\bar u \cdot\ +v D_x^{-1}\re\bar u\cdot\
\\
-\tfrac{1}{2}\i u +u D_x^{-1} v 
&
D_x -\i v +u D_x^{-1}\re\bar u\cdot\
\epm . 
\end{equation}

The pair of operators \eqref{2nd.Hop}--\eqref{2nd.Jop}
provides a bi-Hamiltonian formulation
\begin{equation}
\bpm v_t \\ u_t \epm 
= \Hop\bpm \delta\mk{H}/\delta v \\ \delta\mk{H}/\delta\bar u^\t \epm
= \Eop\bpm \delta\mk{E}/\delta v \\ \delta\mk{E}/\delta\bar u^\t \epm
\end{equation}
where $\Eop= \Rop\Hop$ is Hamiltonian operator compatible with $\Hop$. 
The first Hamiltonian is given by 
$\mk{H}=\int \big( \re( \bar u\cdot(u_x\hatJ) ) + v\im( \bar u\cdot(u\hatJ) \big)\, dx$,
while the second Hamiltonian is similar to the one for the $SU(2)$ equation \eqref{1st.su2.ueqn}. 
Likewise, system \eqref{2nd.su2.ueqn} is not Hamiltonian with respect to $\hatJ$. 

A Lax pair \eqref{zerocurveqn} for system \eqref{2nd.su2.ueqn}
is given by the $\mk{so}(6)$ matrices 
\begin{equation}\label{so.U.V}
U =
\bpm \EE+\re\UU &\im\UU \\ \im\UU & \EE+\re\UU \epm , 
\quad
V =
\bpm \re(\HH+\WW) &\im(\HH+\WW) \\ \im(\HH-\WW) & -\re(\HH-\WW) \epm 
\end{equation}
where
\begin{align}
& 
\EE = 
\bpm 0& \chi & 0 \\ -\chi & 0 & 0 \\ 0 & 0 & 0 \epm,
\quad
\UU  = \bpm \i v & 0 & u_1 \\  0 & \i v & u_2 \\ -u_1  & -u_2  & 0 \epm,
\\
&
\HH = \bpm 0 & 0 & -\chi\re(u\hatJ)_2 \\  0 & 0 & \chi\re(u\hatJ)_1 \\ \chi\re(u\hatJ)_2 & -\chi\re(u\hatJ)_1 & 0 \epm,
\\
&
\WW = 
\bpm 
\i \tfrac{1}{2}\im( \bar u\cdot (u\hatJ) ) + \i w_{1\pa} &  w_{2\pa} & (u_x\hatJ -\i v u\hatJ)_1\\
-\bar w_{2\pa} & \i \tfrac{1}{2}\im( \bar u\cdot (u\hatJ) ) -\i w_{1\pa} & (u_x\hatJ -\i v u\hatJ)_2\\
-(\bar u_x\bar\hatJ +\i v \bar u\bar\hatJ)_1 & -(\bar u_x\bar\hatJ +\i v \bar u\bar\hatJ )_2 & \i \im( \bar u\cdot (u\hatJ) ) 
\epm ,
\end{align}
with
\begin{equation}
\bpm -\i w_{1\pa} & \bar w_{2\pa} \\ -w_{2\pa} & \i w_{1\pa} \epm
= D_x^{-1}( \bar u^\t \cdot u_x \hatJ + \hatJ \bar u_x^\t \cdot u + \i v [\hatJ,\bar u^\t\cdot u] )_0 -\chi^2 \hatJ . 
\end{equation}
The spectral parameter in the Lax pair is $\chi$.

\subsection{Derivation}

The $SU(2)$ system \eqref{2nd.su2.ueqn} comes from 
applying the general zero-curvature framework outlined in section~\ref{sec:nls}
to the symmetric Lie algebra $(\mk{so}(6),\mk{u}(3))$. 
This Lie algebra has the bracket relations \eqref{brack.rel}
with $\hsp = \mk{u}(3)$, $\msp = \Rnum^6 \simeq \mk{so}(6)/\mk{u}(3)$, 
and $\gsp=\mk{so}(6)$. 
Its rank is $1$ \cite{Hel}, 
and so there is a unique choice of $\e$ in a Cartan space in $\msp$
up to gauge freedom generated by $\ad(\hsp)$. 

A matrix representation is given by 
\begin{align}
& \gsp = \bpm \re(\AA+\BB) &\im(\AA+\BB) \\ \im(\AA-\BB) & -\re(\AA-\BB) \epm 
\simeq \mk{so}(6),
\quad 
\re\BB,\re\AA,\im\AA\in\mk{so}(3), \im\BB\in\mk{s}(3)
\label{so.matr.rep}
\\
& \hsp = \bpm \re\BB &\im\BB \\ -\im\BB& \re\BB \epm
\simeq \mk{u}(3), 
\quad 
\BB\in  \mk{u}(3) 
\label{so.hsp.matr.rep}
\\
& \msp =  \bpm \re\AA& \im\AA\\ \im\AA& -\re\AA \epm
\simeq \Cnum^3 \simeq \Rnum^6,
\quad 
\AA\in\mk{so}(3,\Cnum) 
\label{so.msp.matr.rep}
\end{align}
and
\begin{equation}
\e = \bpm \EE & \mb{0} \\ \mb{0} & -\EE \epm
\end{equation}
where $\chi$ is arbitrary non-zero real constant related to norm of $\e$
(namely, $K(\e,\e)=-4\chi^2$). 
Note $\mk{s}(n)$ denotes the space of symmetric $n\times n$ matrices. 

The isotropy group preserving $\e$ is $H_\pa = SU(2)\times U(1)\subset U(3)$,
whose Lie algebra is $\hsp_\pa\simeq\mk{su}(2)\oplus\mk{u}(1)\in\mk{u}(3)$. 
This Lie algebra contains the normalized generator \eqref{hatJ}:
\begin{equation}\label{so.symm.su2}
\hatJ \simeq \bpm \re\BB_\hatJ &\im\BB_\hatJ \\ -\im\BB_\hatJ & \re\BB_\hatJ \epm,
\quad
\BB_\hatJ = 
\bpm 
\i\cos\theta & -e^{-\i\psi}\sin\theta & 0 \\
e^{\i\psi}\sin\theta &-\i\cos\theta & 0\\
0 & 0 & 0 
\epm .
\end{equation}

The orthogonal decompositions 
$\msp=\msp_\pa\oplus\msp_\pe$ and $\hsp=\hsp_\pa\oplus\hsp_\pe$ 
correspond to $\AA = \AA_\pa+\AA_\pe$ and $\BB = \BB_\pa+\BB_\pe$: 
\begin{align}
\AA_\pa &= 
\bpm 0& a_\pa & 0\\  -a_\pa & 0 & 0 \\ 0 & 0 & 0 \epm
\in\mk{so}(3,\Cnum),
\quad
a_\pa\in\Rnum
\label{so.msp.par.matr.rep}
\\
\AA_\pe &= 
\bpm 0& \i a_\pe & a_{1\pe} \\  -\i a_\pe & 0 & a_{2\pe} \\ -a_{1\pe}  & -a_{2\pe}  & 0 \epm
\in\mk{so}(3,\Cnum),
\quad
a_\pe\in\Rnum,
a_{1\pe},a_{2\pe}\in\Cnum 
\label{so.msp.perp.matr.rep}
\\
\BB_\pa &= 
\bpm 
\i b_{1\pa}  & b_{2\pa} & 0 \\
-\bar b_{2\pa} &-\i b_{1\pa} & 0\\
0 & 0 & \i b_\pa
\epm
\in\mk{u}(3),
\quad
 b_{1\pa}, b_\pa\in\Rnum,
 b_{2\pa}\in\Cnum ,
\label{so.hsp.par.matr.rep}
\\
\BB_\pe &= 
\bpm 
\i b_\pe& 0 &  b_{1\pe}\\
0 & \i b_\pe& b_{2\pe}\\
-\bar b_{1\pe} & -\bar b_{2\pe} & 0
\epm
\in\mk{u}(n),
\quad
 b_\pe\in\Rnum,
 b_{1\pe}, b_{2\pe}\in \Cnum . 
\label{so.hsp.perp.matr.rep}
\end{align}
The scaled space $\hsp^\pe:=\tfrac{1}{\chi^2}\ad(\e)\msp_\pe\simeq\hsp_\pe$
is given by the correspondence
\begin{equation}
\BB^\pe := \tfrac{1}{\chi}
\bpm 
-2\i a_\pe& 0 & a_{2\pe}\\
0 & -2\i a_\pe& -a_{1\pe}\\
-\bar a_{2\pe} & \bar a_{1\pe} & 0
\epm .
\end{equation}

The matrices in the zero-curvature equation \eqref{zerocurveqn} in $\mk{so}(6)$ 
have the form \eqref{so.U.V} with 
\begin{align}
& 
\UU  = \bpm \i v & 0 & u_1 \\  0 & \i v & u_2 \\ -u_1  & -u_2  & 0 \epm,
\quad
v\in\Rnum,
u_1,u_2\in\Cnum
\\
&
\HH = \bpm 0 & \chi h_\pa +\i \tfrac{1}{2}\chi H^\pe & -\chi h_2^\pe \\  -\chi h_\pa -\i \tfrac{1}{2}\chi H^\pe & 0 & \chi h_1^\pe \\ \chi h_2^\pe  & -\chi h_1^\pe  & 0 \epm,
%\HH = \bpm 0 & \chi h_\pa +\i h_\pe & h_{1\pe} \\  -\chi h_\pa -\i h_\pe & 0 & h_{2\pe} \\ -h_{1\pe}  & -h_{2\pe}  & 0 \epm
\quad
h_\pa,H^\pe\in\Rnum,
h_1^\pe,h_2^\pe\in\Cnum
\\&
\WW = 
\bpm 
\i W_\pe + \i w_{1\pa} &  w_{2\pa} & w_{1\pe}\\
-\bar w_{2\pa} & \i W_\pe -\i w_{1\pa} & w_{2\pe}\\
-\bar w_{1\pe} & -\bar w_{2\pe} & \i W_\pa
\epm,
\quad
w_{1\pa},W_\pa,W_\pe\in\Rnum,
w_{2\pa},w_{1\pe},w_{2\pe}\in\Cnum
\end{align}
%$h_\pe=\tfrac{1}{2}\chi h^\pe$,
%$(h_{1\pe},h_{2\pe}) = \chi (-h_2^\pe,h_1^\pe)$
%$(H^\pe,h_1^\pe,h_2^\pe) =\tfrac{1}{\chi}(-2h_\pe,h_{2\pe},-h_{1\pe})$

Here $(v,u_1,u_2)$ are the Hasimoto variables, 
and $(H^\pe,h_1^\pe,h_2^\pe)$ are the flow variables associated to the space $\hsp^\pe$
(see \Ref{Anc2008}).
The components of the zero-curvature equation yield the system 
\begin{subequations}\label{so.zerocurv.components}
\begin{align}
& 
D_x h_\pa = v H^\pe+\re( \bar u_1 h_1^\pe + \bar u_2 h_2^\pe ),
\\
&
D_x w_{1\pa} = -\im( \bar u_1 w_{1\pe} - \bar u_2 w_{2\pe} ),
\quad
D_x w_{2\pa} = u_1 \bar w_{2\pe} - \bar u_2 w_{1\pe} ,
\\
&
D_x W_\pa = 2\im( \bar u_1 w_{1\pe} + \bar u_2 w_{2\pe} ),
\\
& 
W_\pe = \tfrac{1}{4}D_x H^\pe + v h_\pa +\tfrac{1}{2}\im( \bar u_1 h_1^\pe +\bar u_2 h_2^\pe ),
\\&
w_{1\pe} = D_x h_1^\pe +u_1 (h_\pa -\tfrac{1}{2}\i H^\pe) -\i v h_1^\pe,
\quad
w_{2\pe} = D_x h_2^\pe +u_2 (h_\pa -\tfrac{1}{2}\i H^\pe) -\i v h_2^\pe,
\\
&
v{}_t = D_x W_\pe +\im( \bar u_1 w_{1\pe} +\bar u_2 w_{2\pe} ) +\chi^2 H^\pe,
\\
&
u_1{}_t = D_x w_{1\pe} +\i v w_{1\pe} -\i u_1( W_\pe -W_\pa+w_{1\pa} ) -u_2 w_{2\pa} +\chi^2 h_1^\pe,
\\
&
u_2{}_t = D_x w_{2\pe} +\i v w_{2\pe} -\i u_2( W_\pe -W_\pa -w_{1\pa} ) +u_1 \bar w_{2\pa} +\chi^2 h_2^\pe . 
\end{align}
\end{subequations}
This system is invariant under an $SU(2)$ symmetry that 
comes from the action of the isotropy group $H_\pa = SU(2)$ 
on $(v,u_1,u_2)$, $(W_\pe,w_{1\pe},w_{2\pe})$, $(H^\pe,h_1^\pe,h_2^\pe)$. 
By going to the 2-component spinor variables \eqref{u.pair}, \eqref{w.hperp.pair},
along with the similar matrix variable
\begin{equation}\label{so.wpar.matr}
\mb{w}_\pa = \bpm -\i w_{1\pa} & \bar w_{2\pa} \\ -w_{2\pa} & \i w_{1\pa} \epm, 
\end{equation}
the system \eqref{so.zerocurv.components} takes the simpler form 
\begin{subequations}\label{so.zerocurv.system}
\begin{align}
& 
D_x h_\pa = v H^\pe +\re(\bar u\cdot h^\pe),
\\
& 
W = \tfrac{1}{4}D_x H^\pe + v h_\pa +\tfrac{1}{2}\im(\bar u \cdot h^\pe),
\\
&
w = D_x h^\pe -\i v h^\pe + u (h_\pa -\tfrac{1}{2}\i H^\pe) ,
\\
& 
D_x W_\pa = 2\im(\bar u\cdot w),
\quad
D_x \mb{w}_\pa = ( \bar u^\t \cdot w - \bar w^\t \cdot u )_0,
\\
& 
v_t = D_x W +\im(\bar u \cdot w) +\chi^2 H^\pe , 
\\
&
u_t = D_x w +\i v w  +\i u (W_\pa -W) + u \mb{w}_\pa +\chi^2 h^\pe , 
\end{align}
\end{subequations}
where, for ease of notation, $W:=W_\pe$. 
Specifically, the normalized $SU(2)$ symmetry generator \eqref{so.symm.su2}
becomes $(v,u)\to (0,u\J)$, which acts trivially on $v$. 

Now, the $SU(2)$ symmetry generator is used to define a flow 
\begin{equation}
H^\pe =0 , 
\quad
h^\pe = u \hatJ . 
\end{equation}
The system \eqref{so.zerocurv.system} then yields 
\begin{gather}
h_\pa = c_1,
\\
W = \tfrac{1}{2}\im( \bar u\cdot (u\hatJ) ) +c_1 v,
\quad
w = u_x\hatJ -\i v u\hatJ +c_1 u , 
\\
W_\pa = \im( \bar u\cdot (u\hatJ) ) + C_1,
\quad
\mb{w}_\pa = D_x^{-1}( \bar u^\t u_x \hatJ + \hatJ \bar u_x^\t u )_0
+ \i D_x^{-1}(v [\hatJ,\bar u^\t u] ) +\CC_2 , 
\end{gather}
and 
\begin{equation}
\begin{aligned}
&\begin{aligned}
v_t = 2\im( \bar u_x\cdot (u\hatJ)) +c_1 v_x  , 
\end{aligned}
\\
&\begin{aligned}
u_t & = 
u_{xx}\hatJ -\i v_x u\hatJ + v^2 u\hatJ 
+\tfrac{1}{2}\i \im( \bar u\cdot (u\hatJ) ) u
\\&\qquad
+ u D_x^{-1}( \bar u^\t u_x \hatJ + \hatJ \bar u_x^\t u )_0 + \i u D_x^{-1}(v [\hatJ,\bar u^\t u] )
%\\&\qquad
+c_1 u_x + u (\i C_1 + \CC_2 +\chi^2 \hatJ) , 
\end{aligned}
\end{aligned}
\end{equation}
where $c_1,C_1$ are arbitrary constants, 
and $\CC_2$ is an arbitrary constant matrix in $\mk{su}(2)$. 
Hence, the $SU(2)$ NLS-type system \eqref{2nd.su2.ueqn}
is obtained for $c_1=C_1=0$ and $\CC_2=-\chi^2\hatJ$. 

The integrability structure is obtained directly from the system \eqref{so.zerocurv.system} 
similarly to that for the previous $SU(2)$ equation.

\section{Geometric flows}\label{sec:geomflows}

A primary geometric formulation of the NLS equation is given by 
the bi-normal flow of an inelastic curve $\vec{r}(x)$ in $\Rnum^3$,
where $x$ is arclength. 
The bi-normal flow equation consists of 
$\vec{r}_t = \vec{r}_x\times \vec{r}_{xx}$. 
Its well-known equivalence to the NLS equation comes from 
the Hasimoto transformation \cite{Has} $u=\kappa e^{\i\int \tau\,dx}$
in terms of the curvature $\kappa$ and torsion $\tau$ of the curve
(see \Ref{MarSanWan,AncMyr} for details). 

For the later generalizations, 
a useful observation is that the bi-normal equation in $\Rnum^3$
can be expressed in the form 
$\vec{r}_t =\Iop_{\vec{r}}(D_x\vec{r}_x)$
with $\Iop_{\vec{r}} = \vec{r}_x\times$ 
being a geometric representation of $U(1)$ symmetry generator $\i$. 
In particular, this operator has the properties 
$\Iop(\vec{r}_x)=0$
and 
$\Iop_{\vec{r}} ^2=-\Pop^\perp_{\vec{r}}$, 
where 
$\Pop^\perp_{\vec{r}}$ projection operator onto normal plane with respect to $\vec{r}_x$. 

The bi-normal equation can also be expressed as
$\vec{r}_t = \kappa\mb{B}$
in terms of a Frenet frame:
$\mb{T}=\vec{r}_x$ is the tangent vector;
$\mb{N} = \tfrac{1}{\kappa} D_x\mb{T}$ is the principal normal vector;
$\mb{B} = \mb{T}\times \mb{N}$ is the bi-normal vector. 
This formulation leads directly to a Schrodinger map equation 
by identifying the unit vector $\mb{T}$ in $\Rnum^3$ 
with a map $\gamma(t,x)$ into $S^2$
(for details, see \Ref{AncMyr}).
The resulting flow is given by 
$\gamma_t = J(\nabla_x\gamma_x)$,
where $\nabla$ is the covariant derivative (Riemannian connection) on $S^2$, 
and where $J$ is the complex structure tensor on $S^2$, 
satisfying $\nabla J=0$. 
Here $\nabla_x:=\gamma_x\hook\nabla$ is the covariant derivative along $\gamma$. 

There is a similar geometric formulation of the $SU(2)$ integrable systems \eqref{1st.su2.ueqn} and \eqref{2nd.su2.ueqn}. 
The Hasimoto transformation is carried out by means of the following identifications:
\begin{equation}
\vec{r}_x \leftrightarrow U_{\msp},
\quad
\vec{r}_t \leftrightarrow V_{\msp_\pe},
\quad
D_x\vec{r}_x \leftrightarrow [U_{\msp},U_{\hsp}]
\end{equation}
where $U$ and $V$ are the matrices in the Lax pair, 
with the subscripts denoting projections. 
These identifications can be derived from the general formulation of 
inelastic curve flows in symmetric spaces presented in \Ref{Anc2008,AncAsa2019}. 
They will produce a $SU(2)$ bi-normal equation,
from which an equivalent $SU(2)$ Schrodinger map equation is readily obtained. 
A key difference compared to the NLS case is that 
the $\mk{u}(1)$ generator given by the complex structure tensor on $S^2$ 
will be replaced by an $\mk{su}(2)$ generator that is attached to normal space of the curve $\gamma$.

\subsection{Geometric $SU(2)$ flow in $\mk{su}(4)/\mk{sp}(2)$}

Consider an inelastic curve $\vec{r}(x)$ in $\Rnum^5\simeq \mk{su}(4)/\mk{sp}(2)$,
where $x$ is the arclength. 
A geometric representation of the $SU(2)$ symmetry generator $\hatJ$
is given by a vector operator $\hatJ_{\vec{r}}$ that has the properties 
$\hatJ_{\vec{r}}(\vec{r}_x)=0$
and 
$\hatJ_{\vec{r}} ^2=-\Pop^\perp_{\vec{r}}$,
where $\Pop^\perp_{\vec{r}}$ is the projection operator onto the 4-dimensional normal plane with respect to $\vec{r}_x$. 
These two properties determine $\hatJ_{\vec{r}}$ up to a sign. 

The $SU(2)$ bi-normal flow is given by 
\begin{equation}\label{su2.binormal.flow}
\vec{r}_t = \hatJ_{\vec{r}}(D_x\vec{r}_x) . 
\end{equation}
An equivalent formulation consists of 
$\vec{r}_t = \kappa\mb{B}$
where $\mb{B} = \hatJ_{\vec{r}}(\mb{N})$ is a bi-normal vector 
given in terms of the principal normal vector $\mb{N} = \tfrac{1}{\kappa} D_x\mb{T}$ and the tangent vector $\mb{T}=\vec{r}_x$,
with $\kappa$ being the curvature. 

The unit vector $\mb{T}$ in $\Rnum^5$ 
can be identified with a map $\gamma(t,x)$ into $S^4$. 
The resulting flow can be shown to have the form of a generalized $SU(2)$ Schrodinger map
\begin{equation}\label{su2.Schr.map.flow}
\gamma_t =  J_\gamma(\nabla_x\gamma_x)
\end{equation}
where 
$\nabla$ is the covariant derivative (Riemannian connection) on $S^4$, 
and where 
$J_\gamma$ is the $\mk{su}(2)$ generator in the normal subspace of the tangent space along $\gamma$ in $S^4$.

\subsection{Geometric $SU(2)$ flow in $\mk{so}(6)/\mk{u}(3)$}

Similarly, consider an inelastic curve $\vec{r}(x)$ in $\Rnum^6 \simeq \mk{so}(6)/\mk{u}(3)$,
where $x$ is arclength. 
Recall that $(\mk{so}(6),\mk{u}(3))$ is a Hermitian symmetric Lie algebra,
and let $\mb{J}\in\mk{u}(3)$ denote the element representing the hermitian structure. 
This element is the generator of a $\mk{u}(1)$ subalgebra in $\mk{u}(3)$,
which acts on $\mk{so}(6)/\mk{u}(3)\simeq\Rnum^6$ via $\ad(\mb{J})$. 

A geometric representation of $SU(2)$ symmetry generator $\hatJ$ is given by 
the operator $\hatJ_{\vec{r}}$ that is determined (up to a sign) by the properties 
$\hatJ_{\vec{r}}(\vec{r}_x)=0$, 
$\hatJ_{\vec{r}}(\ad(\mb{J})\vec{r}_x)=0$,
and $\hatJ_{\vec{r}} ^2=-\Pop^\perp_{\vec{r}^\Cnum}$,
where 
$\Pop^\perp_{\vec{r}^\Cnum}$ is the projection operator 
onto the 4-dimensional plane that is orthogonal to the span of $\vec{r}_x$ and $\ad(\mb{J})\vec{r}_x$.
Namely, this plane is the complexified normal plane defined relative to the complexified tangent space of the curve. 

A remark here is that $\mb{J}$ does not belong to the isotropy Lie subalgebra $\hsp_\pa\simeq\mk{su}(2)\oplus\mk{u}(1)$; 
accordingly, note that the $\mk{u}(1)$ generator in $\hsp_\pa$ does not coincide with the hermitian structure $\mb{J}$. 

Using these structures, 
the resulting $SU(2)$ bi-normal flow has the form \eqref{su2.binormal.flow}.
This flow is equivalent to the $SU(2)$ Schrodinger map equation \eqref{su2.Schr.map.flow}
on $S^5$.

\section{Concluding remarks}
\label{sec:remarks}

The zero-curvature framework used to obtain the two novel 
spinor/quaternion NLS-type systems \eqref{1st.su2.ueqn} and \eqref{2nd.su2.ueqn}
utilizes the structure of the isotropy Lie subalgebra $\hsp_\pa\supset \mk{su}(2)$ 
in a symmetric Lie algebra $(\gsp,\hsp)$ with a choice of a constant element $\e$ 
in the Cartan space $\msp\simeq\gsp/\hsp$. 

For the two symmetric Lie algebras 
$\mk{su}(4)/\mk{sp}(2)\simeq\mk{so}(6)/\mk{so}(5)$ 
and $\mk{so}(6)/\mk{u}(3)$,
there is a unique choice of $\e$ up to the gauge action of $\ad(\hsp_\pa)$. 
The isotropy subalgebra in the second symmetric Lie algebra contains a $\mk{u}(1)$ factor,
and its generator can be used to derive an integrable $U(1)$ NLS-type system. 
This system turns out to be a 2-component version of 
the integrable Yajima-Oikawa (YO) system \cite{YajOik}. 
In contrast, 
the isotropy subalgebra in the first symmetric Lie algebra has no $\mk{u}(1)$ factor,
and so no $U(1)$ NLS-type system can be derived. 

A picture of the relationship holding between these symmetric Lie algebras
and NLS-type systems is summarized in Table~\ref{table:liealgebras.systems}. 

\begin{table}[h!]
\caption{}
\label{table:liealgebras.systems}
\centering
\begin{tabular}{c|c|c|c||l}
symmetric Lie algebra 
& isotropy 
& symmetry 
& Hasimoto
& integrable 
\\
$\gsp/\hsp$
& subalgebra $\hsp_\pa\subset\hsp$
& generator
& variables
& system
\\
\hline
\hline  
$\mk{so}(4)/\mk{so}(3)$
& $\mk{so}(2)\simeq\mk{u}(1)$
& $\i\in \mk{u}(1)$
& $u\in\Cnum$
& NLS equation
\\
\hline
$\mk{so}(6)/\mk{u}(3)$
& $\mk{u}(1)\oplus\mk{su}(2)$
& $\i\in \mk{u}(1)$
& $v\in\Rnum$, $u\in\Cnum\times\Cnum$ %$u_1,u_2\in\Cnum$
& $2$-component
\\
&
&
&
& YO system
\\
\hline
$\mk{so}(6)/\mk{u}(3)$
& $\mk{u}(1)\oplus\mk{su}(2)$
& $\hatJ\in \mk{su}(2)$
& $v\in\Rnum$, $u\in\Cnum\times\Cnum$ %$u_1,u_2\in\Cnum$
& \eqref{2nd.su2.ueqn}
\\
\hline
$\mk{su}(4)/\mk{sp}(2)$
& $\mk{su}(2)\oplus\mk{sp}(1)$
& $\hatJ\in \mk{su}(2)$
& $u\in\Cnum\times\Cnum$ %$u_1,u_2\in\Cnum$
& \eqref{1st.su2.ueqn}
\\
$\simeq\mk{so}(6)/\mk{so}(5)$
& $\simeq\mk{su}(2)\oplus\mk{so}(3)$
& 
& 
&
\\
\hline
\end{tabular}
\end{table}

Multi-component versions of the spinor/quaternion NLS-type systems 
can be obtained straightforwardly by extending the zero-curvature framework to 
the symmetric Lie algebras 
$\mk{su}(2n-2)/\mk{sp}(n-1)$ and $\mk{so}(2n)/\mk{u}(n)$. 
In the resulting systems, 
$u_1$, $u_2$ will become complex vectors with $n-2$ components, 
whereby $(u_1,u_2)$ can be viewed as a vectorial spinor 
or equivalently a vectorial quaternion. 

Other integrable systems can be derived in a similar way with $SU(2)$ replaced by a larger symmetry group. 
An especially interesting case would be the group $SL(2,\Cnum)$ which is associated to spinors/quaternions in $4$-dimensional Minkowski space. 

An interesting questions for future work will be to study the soliton solutions of 
the two new integrable spinor/quaternion systems \eqref{1st.su2.ueqn} and \eqref{2nd.su2.ueqn}. 
Their Lax pairs are the starting point for 
the construction of an inverse scattering transform that can be used analytically 
to solve the initial value problem.

\section*{Acknowledgements}
S.C.A. is supported by an NSERC research grant. 
E.A. thanks the Mathematics \& Statistics Department of Brock University 
for support during the period in which part this work was initiated.

\end{document}